\documentclass[slac_one]{revtex4}
\usepackage{graphicx}
\usepackage{fancyhdr}
\begin{document}

\title{Top Quark Mass from CDF} 

%

\author{Petteri Meht\"al\"a$^{1}$ \\ (on behalf of the CDF collaboration)}
\affiliation{University of Helsinki - Dept of Physics and Helsinki Institute of Physics \\
Gustaf H\"allstr\"omin katu 2a, FI-00014 University of Helsinki - Finland}
%

\begin{abstract}
This letter presents results on the precision measurement of the top quark mass and a combination of the best CDF top mass measurements. A combination by the TevEWWG (Tevatron electroweak working group) of the best top mass results from CDF and D0 in Run 1 and Run 2 of the Tevatron is also presented.  This result is the current world average, and offers an uncertainty less than 1\%.  The new mass value has been included in traditional LEP EWWG fits to precision electroweak data, and implications for the Standard Model Higgs have been derived.
\end{abstract}

\maketitle

\thispagestyle{fancy}


\section{INTRODUCTION} 
The top quark, discovered in 1995 by both CDF and D0 \cite{cdftopdisc, d0topdisc}, is the heaviest known elementary particle.
The mass of the top quark is of particular interest, as it can be used together with
accurate measurement of the W boson mass to limit mass range of the Standard Model (SM) Higgs boson searches.
The precision measurements of the top quark and
W boson mass serve as a consistency check of the SM if the Higgs boson is discovered, 
and they help to answer whether a discovered scalar particle is indeed the SM Higgs boson.
This letter describes measurement of the top quark mass using data up to 3.0 fb$^{-1}$
collected by the CDF collaboration during Run II at the Tevatron.

\section{PRODUCTION AND DECAY}

Top quarks are predominantly produced in pairs at the Tevatron with a cross-section of about 
7 $pb$. According to SM, they decay
into W boson and a b quark with probability of about one. The decay of the W boson
defines the topology of the $t\bar{t}$ event. The dilepton channel, where both W bosons decay
leptonically to an electron or muon and their corresponding neutrino, has a clean signal, but
suffers from small branching fraction (about 5\%) and underconstrained kinematics. When one W boson decays
leptonically and the other one hadronically, into two quarks, the channel is called lepton+jets channel.
This channel has overconstrained kinematics, event though it has one an undetected neutrino and has
branching fraction of about 30\%. The all-hadronic channel, where both W bosons decay hadronically,
does not have undetected neutrino and has large branching fraction of about 44\%, but suffers from large 
QCD multijet background and jet ambiguities.


\section{DILEPTON ANALYSES}
Due to the unconstrained kinematics, top quark mass measurements in the dilepton channel
must integrate over some unknown quantities.Neutrino Weighting Method scans over
the azimuthal angles of the two neutrinos and reconstructs the top mass estimator using kinematic fit.
The solution corresponding to the minimum of the goodness-of-the fit, $\chi^2$ over azimuthal angles is selected. 
Each possible neutrino longitudinal and W boson b-jet pairing solution is weighted by the goodness-of-the fit.
A template is constructed using the weighted top masses, where the weight is given by $e^{-\chi^2/2}$. 
The measurement on the data corresponding to integrated luminosity of $2.8~fb^{-1}$ 
yields ${\rm M_{top}} = 165.1^{+3.4}_{-3.3} ({\rm stat}) \pm 3.1 ({\rm syst})~{\rm GeV}/c^2$ \cite{cdfdilnw}.
A second measurement presented here, uses matrix element method, where the theoretical production and decay
matrix elements are used to construct most probable top quark mass for each event. 
A matrix element method that is applied on the data of $1.9~fb^{-1}$ uses a evolutionary neural network, where the weights and the topology of the network are optimized for the analysis sensitivity by selecting the strong performers of a population of the networks into the successive generations. 
The evolutionary neural network is applied in the selection stage to 
improve the {\it a priori} statistical uncertainty on the top quark mass by 20\%. 
The analysis measures ${\rm M_{top}} = 171.2 \pm 2.7 ({\rm stat}) \pm 2.9 ({\rm syst})~{\rm GeV}/c^2$ \cite{cdfdilmat}.

\section{LEPTON+JETS ANALYSES}

Using a matrix element multivariate analysis technique,
weighting jet-parton assignments using tagging probability and using neural network based event-by-event
discriminant for background rejection a measurement on data corresponding to integrated luminosity of $2.7~fb^{-1}$ yields
${\rm M_{top}} = 172.7 \pm 1.8 ({\rm stat+JES}) \pm 1.2 ({\rm syst})~{\rm GeV}/c^2$.
The measurement is on July 2008 the most precise single measurement for top mass in the world \cite{cdfljmtm}.



The top quark mass can be measured using the decay length of b-jets and lepton transverse energy in the lepton+jets
channel. Both these quantities are roughly linearly proportional to the top mass and the measurement 
has minimal dependence on the jet energy scale, $JES$. The method applied on the data
corresponding to integrated luminosity of $1.9~fb^{-1}$ yields 
${\rm M_{top}} = 175.3 \pm 6.2 ({\rm stat+JES}) \pm 3.0 ({\rm syst})~{\rm GeV}/c^2$ \cite{cdfljlxy}. 
This analysis is statistically limited, but if this analysis is done at the LHC statistics will no longer be an issue. 
Further, since some of the dominant systematics are statistically limited, 
the results of these techniques could well become competitive with conventional top mass analyses, 
and due to their reduced correlation with conventional top measurements they should help reduce
the uncertainty on the world average top mass in a combination. 


\section{ALL-HADRONIC ANALYSES}
The all-hadronic channel is challenging due to large QCD multijet background.
A dedicated event selection is required to increase the signal-to-background ratio to an acceptable level
for the mass measurement. A neural network based events selection is used in CDF \cite{cdfnn}.
Two recent measurements in all-hadronic channel, both using integrated luminosity of $1.9~fb^{-1}$ from CDF
have been performed. Both analyses use {\it in situ} jet energy scale calibration, data-driven
background model to estimate the shape and amount of the background events with
neural network discrimination. The first analysis uses events with six to eight jets, requires
one or two b-jets. It uses top mass and dijet mass templates for top mass determination. The measurement
yields ${\rm M_{top}} = 176.9 \pm 3.8 ({\rm stat+JES}) \pm 1.7 ({\rm syst})~{\rm GeV}/c^2$ \cite{cdfahtmt}.

\begin{figure}[!th]
\centerline{\includegraphics[width=0.30\columnwidth]{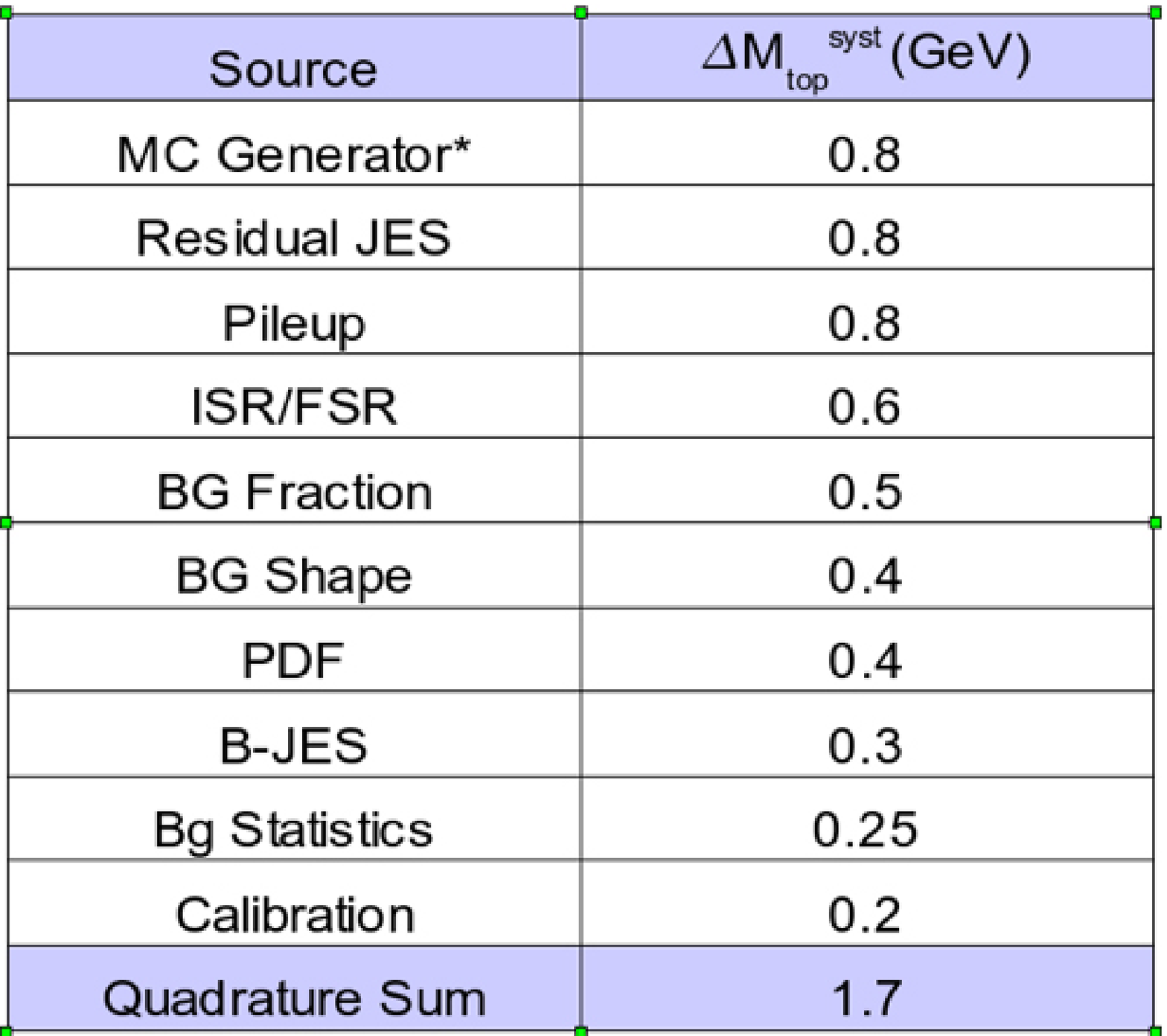}}
\caption{The systematic uncertainties for the Ideogram measurement in the all-hardonic decay channel.}\label{Fig:SYST}
\end{figure}

The second measurement uses Ideogram technique, that uses simplified matrix element for the signal
and templates in top mass and dijet mass for the background and for signal events that have
misreconstruction of the jets. This analysis requires exactly six jets and at least two b-jets
in the event making the signal-to-background fraction much higher, about 2/3. It also measures
the signal fraction from the data simultaneously. 
The measurement yields ${\rm M_{top}} = 165.2 \pm 4.4 ({\rm stat+JES}) \pm 1.7 ({\rm syst})~{\rm GeV}/c^2$ \cite{cdfahideo}.
The individual systematic uncertainties for this analysis are shown in Figure \ref{Fig:SYST}.

\section{TEVATRON COMBINATION}
The combination of D0 and CDF results is performed using BLUE technique \cite{bluemethod}.
In most of the analyses, the jet energy scale is still dominant systematic uncertainty,
but it is expected to decrease as it is measured {\it in situ} when more data will be available
and therefore many analyses still benefit from increased integrated luminosity.
The world average of the best independent measurements of the top quark mass as of March 2008
from D0 and CDF yields  ${\rm M_{top}} = 172.4 \pm 0.7 ({\rm stat}) \pm 1.0 ({\rm syst})~{\rm GeV}/c^2$ and is of 0.7\% accuracy \cite{tevcomb}.

\begin{figure}[t]
\begin{center}
  \begin{tabular}{cc}
    \includegraphics[width=0.35\columnwidth]{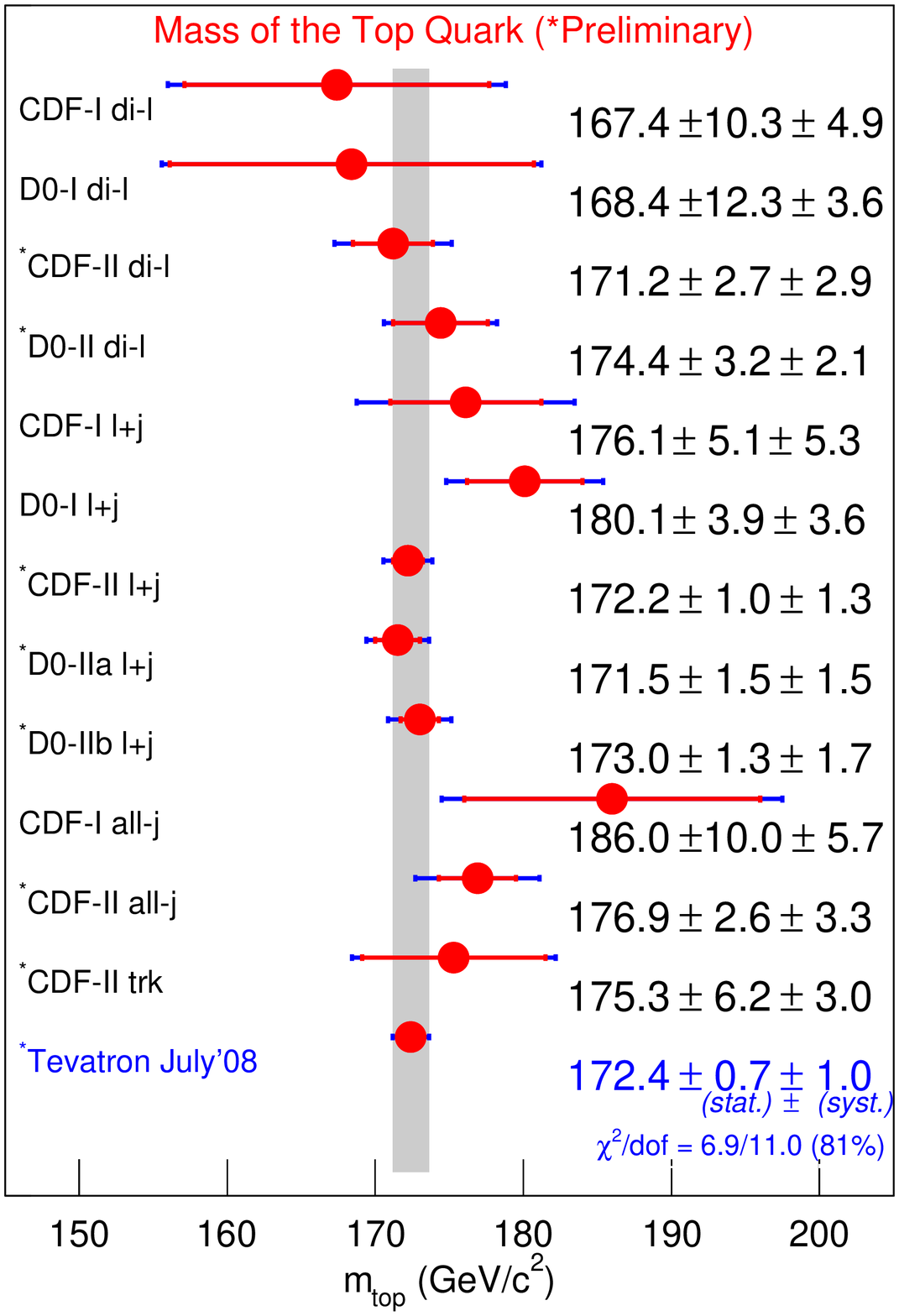}
    \includegraphics[width=0.48\columnwidth]{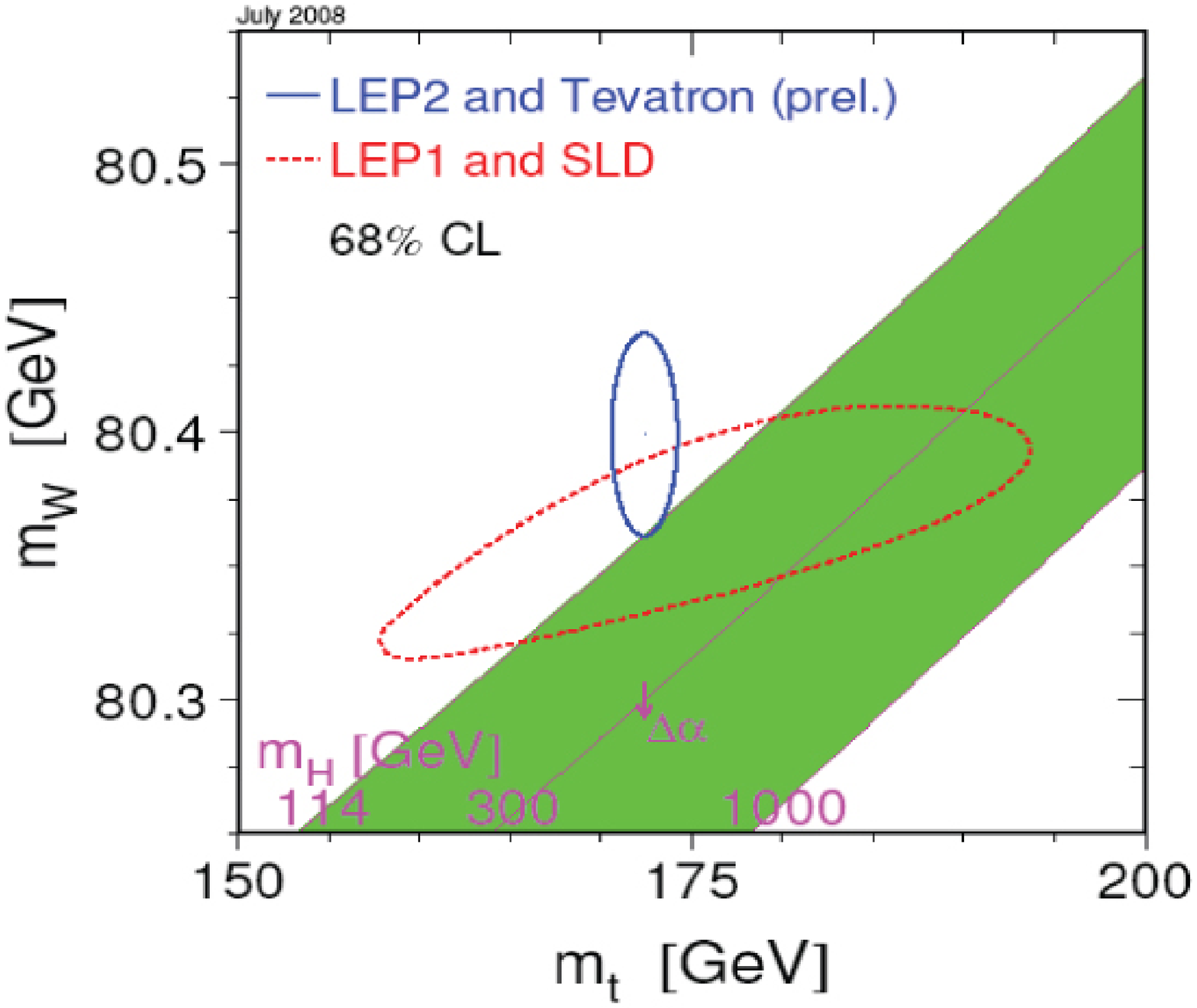}
  \end{tabular}
  \caption{
    The top quark mass world average as on July 2008 (left). The limits on the Standard Model Higgs boson mass as on July 2008 (right).
    \label{Fig:WA}}
\end{center}
\end{figure}

This result can be combined with the current W boson mass measurement to obtain a constraint
for the SM Higgs boson. With 95\% CL SM Higgs boson mass is less than 154$~{\rm GeV}/c^2$.
And the most probable Higgs boson mass value of $m_H = 84^{+34}_{-26} GeV/c^{2}$.
Fig.~\ref{Fig:WA} compares the world average with measurements from both experiments and shows
the current limits on the Standard Model Higgs boson mass in top mass - W boson mass plane.







\end{document}